# From Earthbound to Stars: Analyzing Humanity's Path to a Type II Civilization

Jonathan H. Jiang[1] and Prithwis Das[2]

1. Jet Propulsion Laboratory, California Institute of Technology, Pasadena, CA 91108, USA
2. Department of Computer Engineering, Sejong University, Seoul, South Korea

Correspondence: Jonathan.H.Jiang@jpl.nasa.gov

## Abstract

Humanity is at a critical juncture in its evolution, marked by unprecedented technological advancements and the pursuit of higher civilization statuses as defined by the Kardashev scale. This study provides a comprehensive exploration of humanity's potential progression towards achieving Type I and Type II civilizations, characterized by planetary- and stellar-scale energy utilization, respectively. Building upon Kardashev's framework, we propose refinements that integrate key parameters including energy consumption, information processing, construction mass, and population dynamics. By leveraging machine learning techniques, we analyze global energy data to simulate humanity's energy future, with emphasis on the exponential growth of renewable and nuclear energy sources, and incorporate stellar classifications and insolation flux data from the Planetary Habitability Laboratory to establish energy utilization benchmarks for habitable exoplanets orbiting G-, K-, and M-type stars. Our simulations suggest that humanity could plausibly achieve Type I status by ~2271 CE, enabled by planetary-scale energy harnessing, advanced computational infrastructure, and sustainable population management, while under optimistic assumptions about technological progress and resource utilization, Type II civilization status might emerge between 3200–3500 CE. This projection, however, remains highly contingent on breakthroughs in stellar-scale infrastructures—such as Dyson swarms or Matrioshka Brains—and the sustained integration of interplanetary societies. To more effectively track these trajectories, we introduce a modified Kardashev metric—the Civilization Development Index (CDI)—which balances contributions from energy, information, construction, and population scales, and demonstrate its robustness under varying assumptions. Overall, this study offers a novel, interdisciplinary framework for understanding humanity's long-term trajectory as a multiplanetary civilization, while emphasizing both the promise and uncertainty of forecasting our progression toward stellar-scale futures. Such recognizable existential risks—often described as potential "Great Filters"—could delay, divert, or even prevent this pathway of continued progression, underscoring the urgency of addressing global sustainability and resilience today.

## 1. Introduction

Humanity stands on the precipice of a profound evolutionary leap, propelled by remarkable scientific breakthroughs and technological innovations achieved within an exceptionally short historical timespan. In less than an average human lifespan, we advanced from the Wright brothers' first powered flight to footprints on the lunar surface, epitomizing our species' unparalleled capacity for rapid progress. Yet, as we gaze toward the stars, aspiring toward a future once confined to the realms of science fiction, we are simultaneously confronted by formidable challenges that temper our ambitions. These challenges—ranging from finite natural resources and technological bottlenecks to internal socio-political conflicts—underscore the urgency of deliberate, strategic planning for humanity's long-term survival and advancement.

In this paper, we envision a scientifically rigorous and mathematically grounded pathway by which humanity might navigate these obstacles and ascend the Kardashev scale, a seminal framework developed to categorize civilizations based on their energy consumption capabilities. Introduced in 1964 by Russian astrophysicist Nikolai Kardashev, this classification system has profoundly shaped our understanding of civilization development and inspired decades of interdisciplinary research in fields ranging from astrophysics and astrobiology to sociocultural evolution and futurism. Kardashev delineated three primary civilization types: Type I, which harnesses the total available energy of its home planet; Type II, which utilizes the full energy output of its star; and Type III, capable of tapping into the vast energies of its entire galaxy [1-3].

Since Kardashev's initial proposal, subsequent refinements have enriched the scale's utility. Carl Sagan expanded the original three-category system into a continuous decimal scale to better capture the incremental nature of technological advancement. He also introduced an alphabetic information scale to quantify a civilization's information-processing capabilities, recognizing that energy consumption alone insufficiently characterizes the complexities inherent in technological societies [1, 4]. Further scholarly contributions integrated additional critical dimensions such as large-scale construction mass and population dynamics, reflecting that sustainable civilizational progress involves far more than mere energy utilization [5].

Despite these advancements, a cohesive, integrative model combining these multiple dimensions remains elusive. Our study addresses this gap by introducing modifications to Kardashev's scale, synthesizing recent progress in energy research, information processing, large-scale engineering, and population sustainability. We employ state-of-the-art machine learning algorithms to analyze global energy consumption patterns and to project plausible trajectories for humanity's transition from its current intermediate Type 0 to Type I status toward fully achieving Type I and eventually Type II civilization levels[1]. By incorporating astronomical data from potentially habitable exoplanets and their host stars, we further generalize the Kardashev framework to better reflect universal conditions beyond our solar system.

This comprehensive, interdisciplinary approach seeks not merely to predict humanity's future milestones but to illuminate a scientifically plausible roadmap for achieving them. Our aim is to foster deeper understanding and inspire actionable strategies that will enable our species to responsibly steward technological growth and ecological sustainability. Ultimately, by confronting both the challenges and opportunities inherent in our collective ambition to expand civilization beyond Earth, we aspire to contribute to humanity's enduring quest for meaning and our evolutionary imperative to reach for the stars.

## 2. The Four Scales and Computational Framework

We initiate our analysis by extending and refining the original Kardashev scale, emphasizing four critical dimensions integral to a civilization's advancement: societal information processing capacity (knowledge level), construction mass, population dynamics, and energy utilization. Although extensive literature exists on each dimension individually, an integrative framework elucidating the complex interplay among these factors remains underdeveloped [1]. This study seeks to bridge this gap by creating a comprehensive, unified model that accurately captures the intricate relationships and interdependencies of these dimensions. This integrative model aims to provide reliable estimates for humanity's potential timeline to achieve Type I and eventually Type II civilization statuses, assuming the absence of major global catastrophes or existential threats.

Central to our methodology is the incorporation of recent advancements in computational modeling, particularly leveraging machine learning techniques. These methodologies facilitate refined predictions and detailed scenario analyses of humanity's trajectory along the Kardashev scale. By systematically modeling historical data and forecasting future scenarios, we enhance our understanding of the technological, societal, and ecological milestones required for progressing toward higher civilization statuses.

## 2.1. Energy Scale

Energy utilization is a foundational metric within the Kardashev framework, characterizing a civilization by its ability to harness available energy resources. Traditionally, a Type I civilization is defined as capable of utilizing energy comparable to Earth's total solar insolation, estimated to be between $10^{16}$ and $10^{17}$ watts [1,4]. In this analysis, we extend the conventional energy utilization scale to include exoplanetary systems, thus establishing a universal energy scale applicable beyond Earth. This broader perspective facilitates comparative assessments across potentially habitable exoplanets, providing deeper insights into energy constraints and opportunities at the planetary scale.

To quantitatively characterize potential habitability and energy availability across planetary systems, we utilize insolation flux data from the Potentially Habitable Worlds (PHW) catalog maintained by the Planetary Habitability Laboratory (PHL) at the University of Puerto Rico, Arecibo. Our analysis employs statistical methods, specifically examining the frequency distribution and cumulative distribution function (CDF), as illustrated in Figure 1.

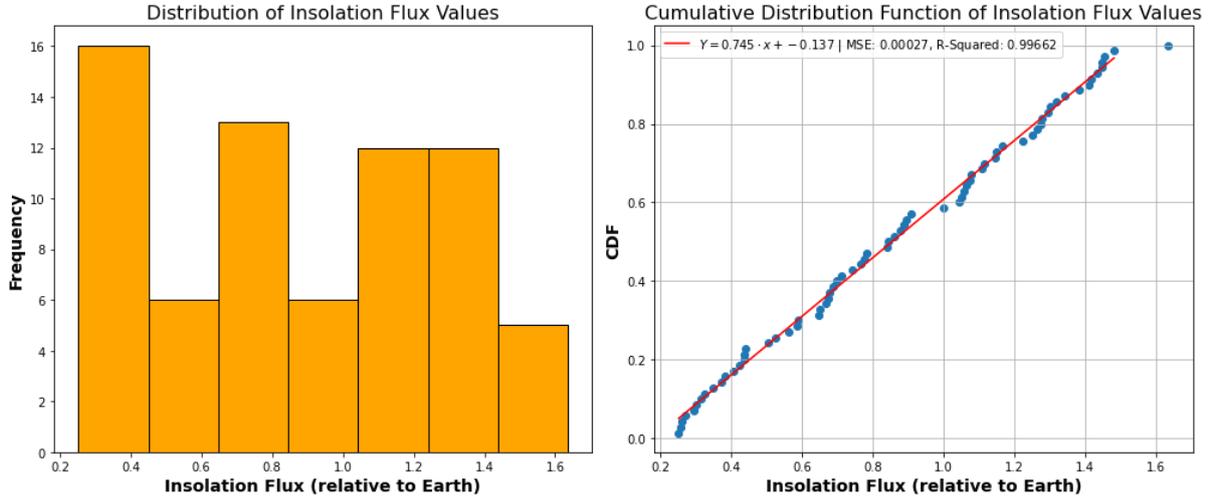

**Figure 1:** (a) Frequency distribution of potentially habitable planets (PHPs) based on insolation fluxes relative to Earth's flux. (b) Cumulative distribution function (CDF) of relative insolation fluxes with fitted linear regression model, achieving a mean squared error (MSE) of 0.00027 and an R-squared value of 0.99662.

This statistical analysis reveals the distribution pattern of insolation fluxes across PHPs. Employing Ordinary Least Squares (OLS) regression within the flux range of 0.25 to 1.6, we derive the following mathematical model:

$$F(S_E) = \begin{cases} m \cdot S_E - b, & \text{for } S_E \in (0.25, 1.6) \\ 1, & \text{for } S_E \geq 1.6 \end{cases} \quad (1)$$

where $m = 0.7450 \pm 0.006$ and $b = 0.1370 \pm 0.006$. The high R-squared value underscores the model's robustness and accuracy.

As a pioneering step toward establishing generalized intervals for energy utilization by Type I civilizations, we analyze the spectral classes of stars associated with PHPs. We calculated the mean insolation flux for these PHPs, filtering the dataset to include only those planets orbiting host stars of specified spectral types, employing advanced data mining techniques consistent with previous studies [6].

Furthermore, we computed the mean insolation flux and its variability for exoplanets orbiting host stars of specific spectral types. The flux variability is captured within one standard deviation from the mean, calculated as:

$$\sigma = \sqrt{\frac{1}{N-1}\sum_{i=1}^{N}(S_{E_i} - S_{E_\mu})^2} \quad (2)$$

where $N$ is the number of PHPs per spectral type, $S_{E_i}$ is the relative flux for the $i^{th}$ exoplanet, and $S_{E_\mu}$ denotes the mean insolation flux for the given spectral type.

Thus, generalized energy utilization intervals for Type I civilizations are defined by stellar spectral class as:

$$Range \coloneqq (S_{E_\mu} - \sigma, S_{E_\mu} + \sigma) \quad (3)$$

Notably, no PHPs are observed orbiting O, B, A, and F-type stars, as depicted in Figure 2. We summarize the resulting energy intervals for G, K, and M-type host stars in Table 1.

A comprehensive evaluation of these energy intervals demonstrates strong consistency with prior research. Notably, the calculated energy interval for planets orbiting G-type stars—approximately $10^{17}$ watts, as initially proposed by Lemarchand [4]—aligns closely with the theoretical power range for Type I civilizations harnessing energy from similar stellar environments, akin to Earth's solar insolation power (~$1.74 \times 10^{17}$ watts).

The energy intervals calculated for K-type and M-type stars are lower, reflecting the reduced luminosity and cooler temperatures characteristic of these star types [7]. Specifically, the energy interval for K-type stars ranges from $1.09 \times 10^{17}$ to $2.17 \times 10^{17}$ watts, whereas the interval for M-type stars spans from $7.16 \times 10^{16}$ to $2.08 \times 10^{17}$ watts. These results underscore the significant impact stellar luminosity has on available energy, further substantiating existing literature.

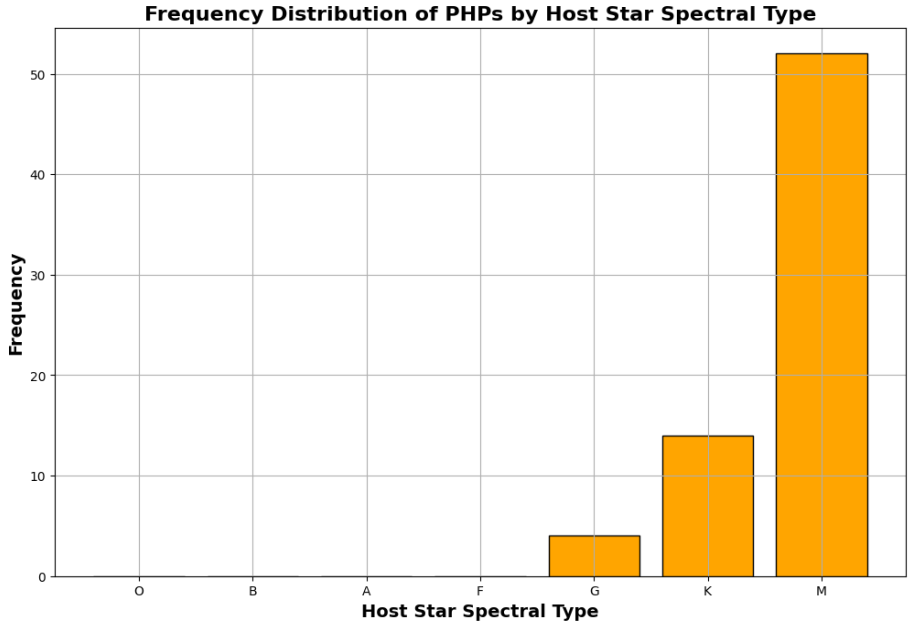

**Figure 2:** Frequency distribution of potentially habitable exoplanets by host star spectral class.

Humanity's transition to a Type I civilization fundamentally depends on its ability to establish a sustainable and resilient energy infrastructure. Projections by the International Energy Agency (IEA) indicate a significant shift from fossil fuels toward renewable sources, estimating that renewables will constitute over 60% of the global energy supply by 2050, alongside substantial growth in nuclear energy capacity [8]. Such energy transitions are critical for meeting the energy demands required to attain Type I civilization status. These historical and projected shifts are illustrated in Figure 3, which shows how renewable energy has recently overtaken nuclear power as the leading clean energy source.

Table 1: Type I Energy Utilization Intervals for Different Host Stars

| Type | Color [7] | Approximate Surface Temperature (in K) [7] | Requisite Energy Interval for attaining Type I State (in Watts) |
|---|---|---|---|
| O | Blue | > 25,000 | - |
| B | Blue | 11,000 – 25,000 | - |
| A | Blue | 7500 – 11,000 | - |
| F | Blue to White | 6000 – 7500 | - |
| G | White to Yellow | 5000 - 6000 | $2.03 \times 10^{17} - 2.78 \times 10^{17}$ |
| K | Orange to Red | 3500 – 5000 | $1.09 \times 10^{17} - 2.17 \times 10^{17}$ |
| M | Red | < 3500 | $7.16 \times 10^{16} - 2.08 \times 10^{17}$ |

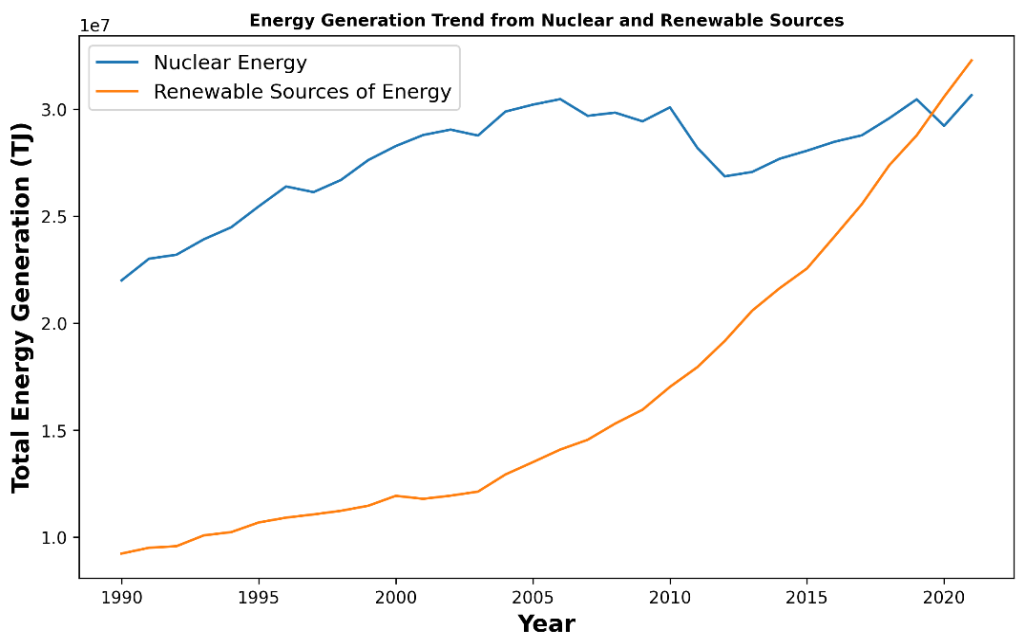

**Figure 3:** Global energy supply trends (1990–2021) from nuclear and renewable energy sources. The data illustrates nuclear energy as historically dominant among clean energy sources; however, recent years have seen renewable energy surpass nuclear power, exhibiting rapid and consistent exponential growth.

Our analysis employs advanced machine learning techniques to model global energy trends, specifically focusing on the growth trajectories of renewable and nuclear energy sources, to simulate the conditions required for achieving Type I civilization status and extrapolating toward Type II conditions. Due to the exponential growth pattern observed in renewable energy adoption, linear regression modeling was applied to logarithmically transformed data, resulting in a robust predictive model, as illustrated in Figure 4.

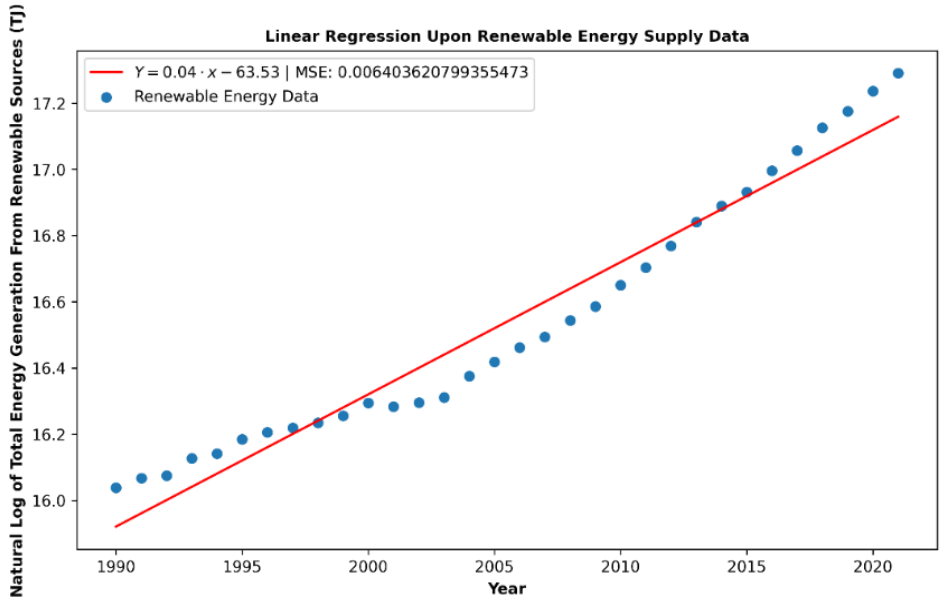

**Figure 4:** Linear regression analysis of renewable energy generation data. The model achieved an R-squared value of approximately 0.96, indicating that it explains 96% of the annual variance in the data, with a mean squared error (MSE) of approximately 0.006.

Conversely, the application of random forest regression to nuclear energy data demonstrated significant model overfitting (MSE: 55,398,692,129.73), rendering it unreliable due to the inherent variability and complexity of nuclear energy data, as depicted in Figure 5. This overfitting underscores the difficulty of capturing long-term nuclear energy trajectories with nonparametric machine learning approaches.

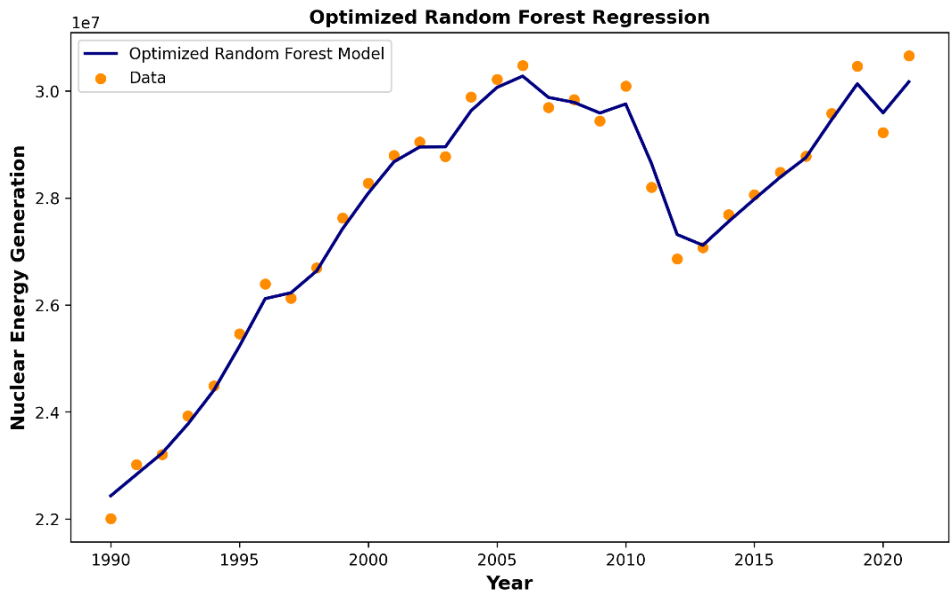

**Figure 5:** Random forest regression analysis of nuclear energy generation data. The model exhibited severe overfitting, indicated by an exceptionally high MSE, making it unsuitable for reliable future projections.

To mitigate these modeling limitations, we adopted a classical exponential growth model for nuclear energy data, assuming a consistent growth rate. This method, widely supported in existing literature, utilizes the Compound Annual Growth Rate (CAGR) to smooth fluctuations over time. Our analysis resulted in a calculated nuclear energy CAGR of 1.04%, slightly below the projected market growth rate of 1.57% for the year 2029 [9].

Incorporating these refined methodologies, we conducted simulations exploring four distinct scenarios. The results are summarized in Table 2. Since although the threshold increases by an order of 10, the average growth rate is also higher. Therefore, the attainment is earlier.

Table 2: Results from Simulation Run for Earth

| Type I Threshold (Watts) | Average Annual Growth Rate for Nuclear Energy (%) | Simulation Results Attainment Year (CE) |
|---|---|---|
| $10^{16}$ | 1.04 | Mean: 2943.8469<br>Range: 2932 - 2944 |
| $10^{16}$ | 1.57 | Mean: 2635.8371<br>Range: 2627 - 2636 |
| $2.03 \times 10^{17}$ | 1.04 | Mean: 3234.9959<br>Range: 3234 - 3235 |
| $2.03 \times 10^{17}$ | 1.57 | Mean: 2271.0<br>Range: 2271 - 2271 |

Synthesizing the refined Type I threshold for habitable planets orbiting G-type stars with projected nuclear and renewable energy growth rates, we predict humanity could achieve Type I civilization status by approximately the year 2271 CE. This estimate closely aligns with previously published literature forecasts, which anticipate attainment between the years 2333 and 2404 CE [10].

To further refine and quantify humanity's technological progression, we propose an enhanced formulation of the Kardashev metric, specifically focused on energy consumption. Our modification integrates observed data and projected growth rates, represented mathematically as:

$$P_Y = N_{Y_0}(1 + CAGR_N)^{Y-Y_0} + e^{(0.0399 \pm 0.002) \cdot Y - (63.5285 \pm 3.173) + \ln\frac{10^9}{31536}} + F(Y_0, Y) \quad (4)$$

Here, $F(Y_0, Y)$ includes all supplementary energy sources, ranging from traditional fossil fuels to speculative cosmic-scale infrastructures. The variables $Y_0$ and $Y$ denote the initial year and the target year of interest, respectively, providing the temporal boundaries for the analysis of energy growth and civilization advancement. By substituting this comprehensive expression into the Kardashev metric, we derive a novel equation for quantifying civilization advancement:

$$K_{P|Y} = \frac{\log\left(N_{Y_0}(1+CAGR_N)^{Y-Y_0} + e^{(0.0399 \pm 0.002) \cdot Y - (63.5285 \pm 3.173) + \ln\frac{10^9}{31536}} + F(Y_0,Y)\right) - 6}{10} \quad (5)$$

Although our refined energy scale significantly enhances the estimation of civilization advancement on the Kardashev scale, we have not yet established a comprehensive theoretical foundation for projecting humanity's transition to Type II civilization. This gap primarily arises from the current absence of reliable data projections associated with transformative technological breakthroughs. To address this challenge, future research must holistically integrate energy consumption models with additional critical parameters, including information processing capabilities, large-scale construction capacities, and societal organization and coordination. The next phase of our research will explore these additional dimensions to fully assess the potential pathways and barriers to humanity's evolution toward Type II status.

## 2.2. Information Scale

In addition to energy consumption, the capability of a civilization to process, store, and transmit information is crucial for its advancement. This capacity can be quantitatively measured by the total amount of data that a civilization can store and process, as well as the speed at which it can transmit this data. Currently, global data generation and processing capacities are measured in zettabytes ($10^{21}$ bytes), experiencing exponential growth driven by advances in computing power, artificial intelligence, and telecommunications. Since the advent of automation, humanity's technological progress has increasingly relied on the efficiency of computing, leading to the development of Figures of Merit (FOMs)—scalar quantities, either dimensionless or with specific units, that measure how effectively a technology performs and its societal value [11,12]. In contrast to capital efficiency measures in computing, the Information Scale—as originally proposed by Kardashev and subsequently refined by Sagan—emphasizes a civilization's capacity to acquire, process, and utilize knowledge. This approach necessitates establishing a standardized Functional Performance Metric (FPM), designed specifically to quantify technological and societal development.

Advances in computer science have progressively approached the theoretical limits of human intelligence, particularly evidenced by tracking annual increases in supercomputers' computational capacities relative to estimates required for human brain emulation, as outlined by Sandberg and Bostrom in 2008 [13,14]. This exponential trajectory could potentially culminate in the realization of mega-scale, super-intelligent computational constructs—initially termed "Matrioshka Brains" (MB) by Robert Bradbury [14]. Such structures could theoretically harness the entire solar energy output (~$10^{26}$ watts), consume significant portions of available solar system resources, and operate at computational limits dictated by fundamental physical laws.

The primary challenge of constructing Matrioshka Brains involves material stability under extreme thermal conditions. Three abundant candidate materials—diamond (thermally stable up to ~1275 K), aluminum oxide (melting point ~2345 K), and titanium carbide (melting point ~3143 K)—have been identified as viable structural components. Operational temperatures for these materials are expected to be maintained within 50–80% of their maximum thermal limits to balance material strength, thermal expansion considerations, and structural integrity. At operational temperatures below approximately 1200 K, emitted thermal radiation primarily consists of low-energy infrared photons (< 0.5 eV), whose direct conversion into electricity is currently technologically inefficient. A feasible alternative approach involves concentrating thermal radiation using reflective mirrors and employing heat engines operating near Carnot efficiency limits to generate electricity for the MB's outer layers.

Incorporating Landauer's Principle provides further insights into computational energy efficiency. This principle sets a theoretical lower bound on the energy required to erase one bit of information [15], expressed as:

$$E \geq k_B T \ln 2 \tag{6}$$

where, $k_B$ is the Boltzmann constant and T is the operational temperature.

Applying this principle, we evaluated the minimum energy thresholds for computations based on the selected structural materials (diamond, aluminum oxide, titanium carbide), considering an operating temperature at approximately 80% of their maximum thermal tolerance, summarized in Table 3.

Notably, fracture toughness data for aluminum oxide were sourced from Direct Metal Oxidation (DIMOX) rather than Gas Pressure Metal Infiltration (GPMI), as the former is more relevant for assessing resistance to micro-cracks due to thermal cycling and micro-impacts in space environments.

**Table 3: Potential MB Structural Components Data**

| Structural Component* | Moh's Hardness | Plane Strain Fracture Toughness $K_{IC}$ (MPa-m$^{1/2}$) | Energy Threshold $E_L$ (J/bit) |
|---|---|---|---|
| Diamond | 10 | 6.5 [16] | ~ $9.76 \times 10^{-21}$ |
| Aluminum Oxide | 9 [19] | 9.5 [17] | ~ $1.795 \times 10^{-20}$ |
| Titanium Carbide | 9-9.5 [18] | - | ~ $2.4063 \times 10^{-20}$ |

* Datasheet-level thermophysical properties for alumina used in our thresholds are taken from manufacturer technical data to complement peer-reviewed toughness measurements [19].

Assuming an idealized scenario of complete power utilization efficiency, we calculated the theoretical upper bounds for computational capacities (bits per second) of Matrioshka Brains composed of each structural material, as illustrated by Equation 7 and summarized in Table 4.

$$Compute\ Capacity\ \left(\frac{bits}{s}\right) \leq \frac{Total\ Solar\ Output\ (W)}{E_L\ (J/bit)} \quad (7)$$

**Table 4: Analyzing Compute of the MBs**

| Structural Component | Compute (bits/s) Upper Bound | Annual Bits Processed Upper Bound |
|---|---|---|
| Diamond | $3.95 \times 10^{46}$ | $1.2471 \times 10^{54}$ |
| Aluminum Oxide | $2.15 \times 10^{46}$ | $6.7804 \times 10^{53}$ |
| Titanium Carbide | $1.60 \times 10^{46}$ | $5.0588 \times 10^{53}$ |

Upon further examination of the compute capacity by evaluating the upper bound of annual bits processed, we conclude that the new Type II information scale threshold is approximately $10^{54}$ bits processed annually. This finding is significant, as surpassing this threshold would signify humanity's transition from a Type II to a Type III civilization. We utilize this threshold by comparing it with annual data processed from 2010 to 2025, as summarized in Table 5.

Applying linear regression analysis to the logarithmic transformation of these data points (Figure 6), our model projects that humanity will cross the theoretical Type II information threshold by approximately the year 2274. This projection closely parallels other literature-based forecasts, which estimate Matrioshka Brain (MB) realization around 2250, informed by silicon wafer production trends and microprocessor manufacturing advancements.

Rather than extending the data-driven analysis, we pivot to a broader discussion of potential future trajectories, emphasizing the role of consciousness in human evolution. Drawing from numerous sources, consciousness is considered an emergent property arising from the organizational complexity and computational efficiency inherent in the human[2] brain. Simplifying this concept within the context of humanity's potential progression toward a Type II civilization and beyond, we reference the theory proposed by visionary scientist Robert A. Freitas. Freitas posits that information processing is the fundamental characteristic of all intelligent systems, with matter-energy serving as the fundamental building blocks.

Consequently, it can be inferred logically that more intelligent life forms have a higher likelihood of survival compared to less intelligent ones, suggesting an evolutionary trend toward increased intelligence [21].

Table 5: Annual Data Processed in Bytes from 2010-2025 [20]

| Year | Annual Volume of Data Processed (Zettabytes) |
|---|---|
| 2010 | 2 |
| 2011 | 5 |
| 2012 | 6.5 |
| 2013 | 9 |
| 2014 | 12.5 |
| 2015 | 15.5 |
| 2016 | 18 |
| 2017 | 26 |
| 2018 | 33 |
| 2019 | 41 |
| 2020 | 64.2 |
| 2021 | 79 |
| 2022 | 97 |
| 2023 | 120 |
| 2024 | 147 |
| 2025 (up to 08/29/2025) | 181 |

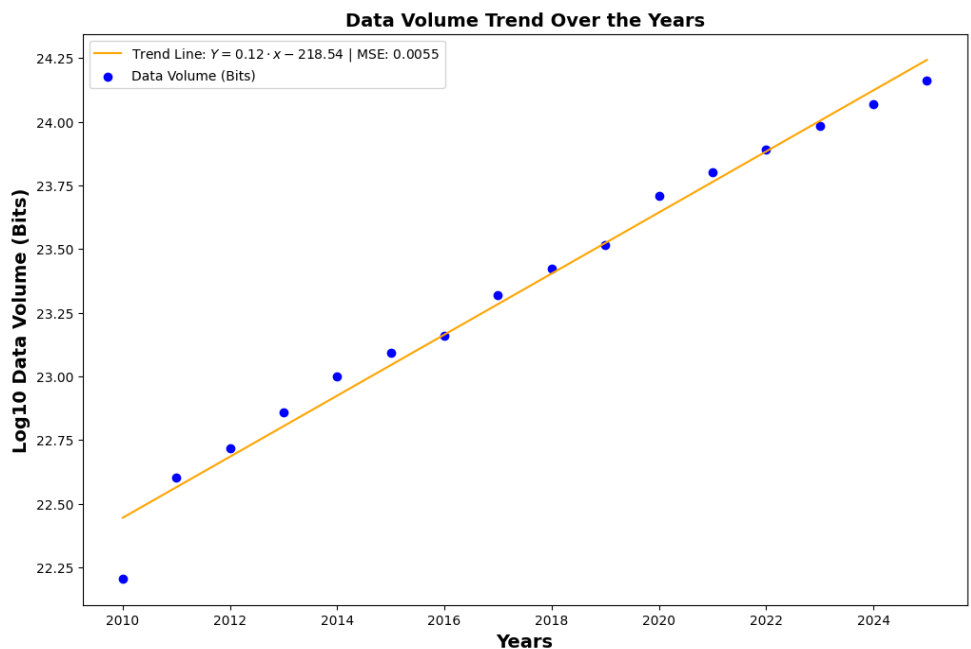

**Figure 6:** Linear Regression applied on Annual Volume of Data Processed. The model achieved an R-Squared value of approximately 0.982, indicating that the model explains 98.2% of the variance present in the data on an annual basis, with the MSE being approximately 0.0055.

Expanding upon this foundation, we employ Freitas's concept of the Sentient Quotient (SQ), which quantifies the cosmic level of sentience applicable to any intelligent entity. The SQ considers both the maximum bit rate processed per unit time (measured in bits per second) and the mass-energy required for processing this data (measured in kilograms). To provide a rigorous mathematical basis for this proposition, we reference the theoretical maximum Sentient Quotient proposed by H.J. Bremermann. Bremermann asserted that if energy states are utilized as information markers, representing the most efficient conceivable markers, then the theoretical maximum volume of information processed by an ideal mass-energy configuration would be approximately $1.4 \times 10^{50}$ bits/sec-kg [23]. This value represents the theoretical upper bound for computational efficiency and intelligence.

Generalizing this concept implies that, within any given mass-energy category, an entity achieving this maximum would possess the highest possible intelligence level in the universe. Such an interpretation could intersect with philosophical or religious frameworks, potentially aligning with the concept of a supreme intelligence or "God." Based on Freitas's estimates, the human brain consists of approximately $10^{10}$ neurons, each with roughly 1000 interconnections and an estimated bit-processing rate of about 1000 bits/sec per neuron. Considering the significant redundancy in the human brain, estimated at a ratio of 1:10,000, and assuming an average human body mass of about 100 kg, the resulting sentience value is approximately $10^{10}$ bits/sec-kg, yielding a Sentient Quotient (SQ) value of approximately 10.

Following this order-of-magnitude principle, Freitas proposed a hierarchical framework consisting of six distinct levels of living consciousness, as summarized in Table 6:

**Table 6: Universal Scale of Sentient Emergents [23]**

| Sentience Level of Individual Entity | Sentience Quotient (SQ) (bits/sec-kg) |
|---|---|
| *Reactivity* | 0 |
| *Consciousness (Present Human State)* | 10 |
| *Communality* | 20 |
| *Hypersociality* | 30 |
| *Galacticity* | 40 |
| *Universality* | 50 |

Consistent with the notion of a communal society representing a subsequent evolutionary stage for humanity, we propose "Communality" as a qualitative indicator of a Type I civilization. Leveraging earlier computations of the upper bound for computational capacity (in bits/s), we normalize these values across component categories by the estimated mass of a Matrioshka Brain (approximately $10^{21}$ kg, comparable to the mass of the asteroid belt [21]). This normalization results in an SQ upper bound of approximately 25 bits/sec-kg, indicating that the transition to a Type II civilization requires achieving a consciousness level situated between Communality and Hypersociality. Collectively, these insights highlight that humanity's progression toward higher civilization states will necessitate not only technological mastery but also significant transformations in the organization, distribution, and collective expression of intelligence.

## 2.3. Large-Scale Construction Mass

Another critical factor in advancing a civilization along the Kardashev scale is its capacity to generate and manage large-scale construction mass. The Construction Mass Scale quantifies a civilization's ability to build, manipulate, and maintain substantial infrastructure required for energy harnessing, resource extraction, transportation, and societal support. As civilizations evolve, their infrastructure projects grow exponentially in scale and complexity, particularly when progressing from planetary to interplanetary and, eventually, interstellar environments. This section examines how advanced construction capabilities underpin a civilization's journey toward Type I and Type II statuses.

### 2.3.1. The Role of Construction Mass in Civilization Advancement

Transitioning to a Type I civilization, which harnesses all available energy resources of its home planet, necessitates extensive infrastructure development. Essential structures include large-scale renewable energy facilities, nuclear fusion reactors, global energy distribution grids, comprehensive transportation networks, resource extraction installations, and orbital habitats or stations. Advancing toward Type II civilization status, where a civilization taps into the energy output of its star system, escalates the scale and complexity of construction considerably.

An iconic example of the immense infrastructure required for a Type II civilization is the Dyson Sphere, a theoretical megastructure that encloses or significantly surrounds a star to capture its energy output. Originally proposed by physicist Freeman Dyson, this concept exemplifies the apex of engineering, demanding extraordinary quantities of construction material, advanced materials science breakthroughs, and highly sophisticated autonomous assembly technologies. A practical variant, the Dyson Swarm—comprising a very large but finite number of independently orbiting satellites—may represent a more feasible engineering solution, still enabling of extensive stellar energy capture and matching the structural inevitability arguments developed for supercivilizations [22,23].

Building such structures would involve sourcing vast quantities of materials from asteroids, moons, and other celestial bodies. These undertakings would also rely on robotic and autonomous technologies, enabling continuous operation over long durations in extreme conditions. Successfully completing a Dyson-like structure would signify a civilization's dominance in stellar-scale energy generation and infrastructure management, firmly establishing its Type II status.

### 2.3.2. Advances in Construction Technologies

Technological innovations play a pivotal role in facilitating large-scale construction projects. Modern advances in additive manufacturing (3D printing), autonomous robotics, artificial intelligence (AI), nanotechnology, and modular construction methods are dramatically reshaping infrastructure development processes. These technologies enable rapid, efficient, and precise construction, significantly reducing dependence on manual human labor and enabling scalable infrastructure assembly.

Autonomous construction systems, guided by advanced AI, would become essential for realizing projects on the scale of a Dyson Sphere or similar stellar megastructures. These robotic systems, potentially capable of self-replication and continuous, unsupervised operation, would extract and process raw materials directly from space-based resources. The exponential scalability enabled by autonomous construction technology represents a critical step in humanity's trajectory toward Type II civilization.

### 2.3.3. The Construction Mass Scale

To quantitatively assess a civilization's large-scale construction capabilities, we introduce the Construction Mass Scale, measuring the cumulative mass of infrastructure constructed over time. This metric captures the ability of civilizations to develop and sustain physical systems required for energy capture, societal advancement, and technological innovation. Mathematically, construction mass at time $t$ can be modeled as:

$$M_c(t) = M_0 \cdot (1 + r)^{t-t_0} \quad (8)$$

where $M_0$ represents initial construction mass, $r$ denotes the construction growth rate, and $t_0$ and $t$ are the initial and current time, respectively. The growth rate $r$ is influenced significantly by technological progress and the capability to exploit planetary and extraplanetary resources.

As civilization progresses from local scales to planetary, interplanetary, and eventually interstellar scales, required infrastructure mass increases exponentially. Precursors to stellar-scale megastructures—such as large-scale space habitats, orbital colonies, lunar and Martian bases, asteroid mining installations, and extensive energy-harvesting arrays—represent stepping stones toward the construction mass necessary for megaprojects like Dyson Spheres.

### 2.3.4. Resource Utilization and Material Science

Large-scale infrastructure projects intrinsically depend on efficient resource utilization and advanced materials science. Civilizations advancing toward Type I and Type II statuses must leverage planetary and off-world resources effectively, including metals, rare minerals, and engineered synthetic materials sourced from asteroids, moons, and planets within their star system.

Breakthroughs in materials science, including graphene, carbon nanotubes, advanced alloys, and nanocomposites offer substantial potential for constructing low mass density yet resilient megastructures capable of enduring extreme environmental conditions in space. For instance, such materials would facilitate revolutionary infrastructure projects like space elevators, orbital rings, and Dyson Swarms. Thus, mastering materials science will critically determine a civilization's capability to undertake and sustain planetary- and stellar-scale, construction projects.

### 2.3.5. Measuring Construction Mass

Analogous to the Kardashev energy scale, we propose a logarithmic Construction Mass Index $C_M$ to assess a civilization's infrastructure growth:

$$C_M = \log_{10}\left(\frac{M_c(t)}{M_{Earth}}\right) \quad (9)$$

where $M_c(t)$ denotes total construction mass at a given time and $M_{Earth}$ represents Earth's current baseline construction mass. This index facilitates comparison across civilizations, highlighting progression toward planetary-scale and eventually stellar-scale infrastructure capabilities.

To provide context, we approximate $M_{Earth}$ as the total mass of constructed human infrastructure, including buildings, roads, railways, dams, bridges, and manufactured materials. Recent estimates suggest this value is on the order of $1.1 \times 10^{12}$ metric tons (~$1.1 \times 10^{15}$ kg) as of the early 21st century, comparable to the biomass of all living organisms on Earth.

Using historical data on global infrastructure and urban development, we can approximate $M_c(t)$ and the resultant $C_M$ for key reference years (Table 7 and Figure 7).

Humanity's ongoing expansion of infrastructure—particularly through space exploration, asteroid mining, and off-Earth industrialization—is expected to significantly elevate this index over the coming centuries. Achieving megastructure-scale projects such as Dyson swarms or orbital habitats would increase construction mass by orders of magnitude, driving CMC_MCM values far above 5. Such a transformation would mark the apex of infrastructure sophistication and represent a critical milestone on the journey toward a Type II civilization.

Table 7: Estimated $M_c(t)$ and the resultant $C_M$

| Year | Estimated Construction $M_c(t)$ (kg) | $C_M = \log_{10}(M_c(t)/M_{Earth})$ | Notes |
|---|---|---|---|
| 1800 | ~1 × 10¹² | -3 | Pre-industrial era, limited infrastructure |
| 1900 | ~1 × 10¹³ | -2 | Industrialization, urban expansion begins |
| 2000 | ~1 × 10¹⁵ | 0 | Modern era, baseline set at $M_{Earth}$ |
| 2100* | ~1 × 10¹⁷ | 2 | Projected growth: megacities, renewable infrastructure, early space industry |

*Projection based on continued exponential growth in infrastructure and early off-Earth resource utilization.

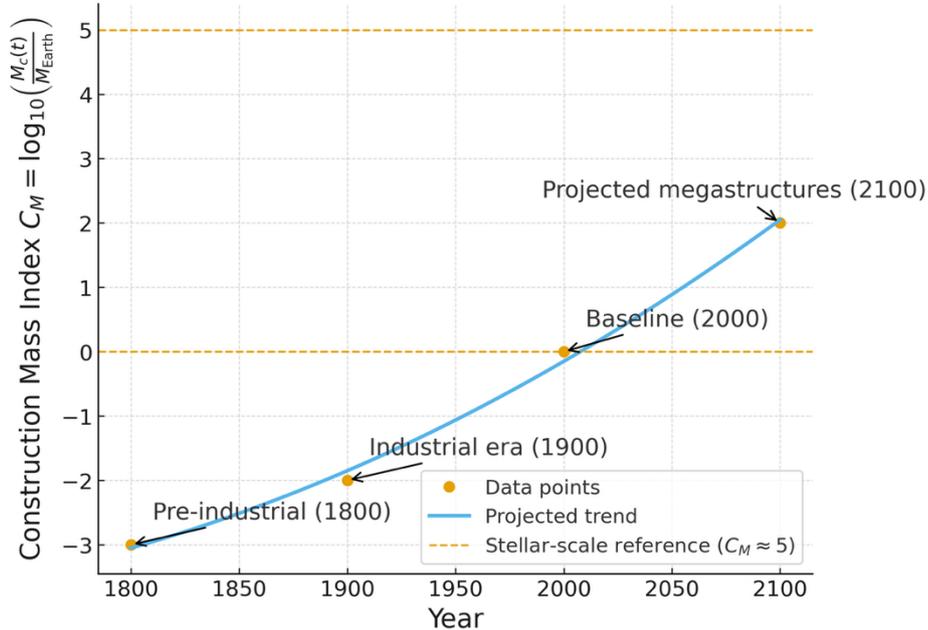

**Figure 7.** Projected trajectory of the Construction Mass Index $C_M$ from 1800 to 2100. Historical estimates (1800–2000) reflect global infrastructure growth; the 2100 value is a projection incorporating megacities, renewable infrastructure, and early space-based construction. Values $C_M \gtrsim 5$ would imply stellar-scale megastructures (e.g., Dyson swarms).

### 2.3.6. Future Projections and the Role of Automation

The future of large-scale construction undoubtedly lies in automation, robotics, and artificial intelligence. Projects of unprecedented complexity and scale will become achievable with minimal human oversight as autonomous robots execute construction, resource extraction, and infrastructure maintenance tasks in space environments. AI-driven technologies will significantly enhance efficiency, scalability, and sustainability in interplanetary and interstellar construction endeavors.

Simultaneously, advances in sustainable materials science, energy-efficient manufacturing processes, and modular assembly technologies will support infrastructure that can withstand harsh space environments. As these technologies mature, theoretical projects such as Dyson Swarms will transition from speculative concepts to feasible engineering endeavors. The completion of such megaprojects would symbolize humanity's transition from planetary to stellar-scale infrastructure capabilities, marking a decisive step toward becoming a Type II civilization.

Ultimately, the Construction Mass Scale offers a vital benchmark for assessing civilizational development. Expanded construction capabilities reflect increasing technological mastery and the capacity for energy harnessing and resource management on the planetary to interstellar scales. The realization of megastructures like Dyson Spheres exemplifies the zenith of large-scale engineering and civilization advancement, profoundly shaping humanity's interstellar future.

## 2.4. Potential for Population Growth

Population growth potential is a critical driver of civilizational advancement, exerting significant influence on societal complexity, technological progression, and resource requirements. As civilizations move toward Type I and eventually Type II statuses, they experience profound transformations in population scale and distribution, shaped by advances in healthcare, technological infrastructure, energy availability, and the exploration and settlement of new environments beyond their planet of origin. This section discusses how population dynamics impact civilizational development and examines factors influencing the sustainable management of expanding populations across planetary, interplanetary, and eventually interstellar scales.

### 2.4.1. Population Growth and Civilizational Expansion

The size, density, and distribution of populations are intrinsically linked to a civilization's technological capabilities and energy infrastructure. A growing population inherently demands increased access to essential resources such as energy, food, water, and shelter, stimulating the advancement of infrastructure necessary to fulfill these needs. For civilizations transitioning toward Type I status, which entails harnessing nearly all available energy on their home planet, expanding populations can intensify pressures on planetary ecosystems and existing infrastructure, prompting a concerted focus on efficiency and sustainable practices.

However, population expansion also serves as a powerful catalyst for innovation, enabling a diverse pool of skills, ideas, and specialized expertise that accelerates technological breakthroughs and enhances societal resilience. This synergistic relationship between population growth and technological advancement supports increasingly sophisticated methods of resource extraction, energy utilization, and environmental management, fostering conditions necessary for fully exploiting planetary resources and eventually extending civilization's reach beyond its home world.

As civilizations advance toward Type II status, characterized by the capacity to harness the energy of an entire star system, population growth must inevitably transcend planetary boundaries. Colonization of extraterrestrial environments, including moons, planetary bodies, and artificial habitats throughout the home star system, becomes essential. This spatial expansion not only addresses the limitations imposed by a single planetary environment but also opens new frontiers for resource acquisition, energy generation, and large-scale construction projects, such as megastructures, which can further facilitate sustainable population growth – i.e., a "virtuous cycle" of robust expansion.

### 2.4.2. Carrying Capacity and Sustainable Growth

The concept of carrying capacity—the maximum population an environment can sustainably support without degradation—is pivotal to understanding and managing population growth. For civilizations approaching Type I status, the carrying capacity is directly influenced by technological advancements in energy production, agricultural efficiency, waste management, and environmental protection. Sustainable growth strategies become critical as populations near planetary limits, necessitating innovations such as vertical agriculture, renewable energy technologies, advanced water purification systems, and comprehensive recycling processes. These technologies collectively enable civilizations to elevate their planetary carrying capacity, ensuring quality of life improvements while minimizing ecological impacts.

For civilizations approaching Type II status, the effective carrying capacity extends beyond planetary confines to encompass an entire star system. Colonizing planetary bodies and establishing artificial habitats significantly expands potential living spaces, facilitating a dynamic carrying capacity that evolves in concert with technological progression and spatial expansion. Notably, large-scale energy infrastructures such as Dyson spheres or swarms become instrumental in supplying abundant energy resources necessary for sustaining extensive multi-interplanetary populations. Fully extending to sustainable interstellar population growth will depend critically on the integration of automated resource management systems, efficient energy distribution networks, and resilient life-support infrastructures.

### 2.4.3. Population Growth Models

Mathematical modeling offers valuable tools for analyzing population dynamics as civilizations advance. A commonly employed approach is the logistic growth model, which captures how populations initially grow rapidly but eventually stabilize as they approach environmental limits:

$$P(t) = \frac{P_{max}}{1 + \left(\frac{P_{max}}{P_0} - 1\right) e^{-r(t-t_0)}} \qquad (10)$$

where:
- $P(t)$: Population size at time $t$
- $P_0$: Initial population size at time $t_0$
- $P_{max}$: Environmental carrying capacity
- $r$: Intrinsic growth rate
- $t_0$: Initial reference time

This model highlights that population growth, while initially exponential, inevitably stabilizes as finite resources impose natural constraints. In the context of a Type I civilization, the carrying capacity is determined by a balance between technological advances and available resources. For Type II civilizations, carrying capacity extends to multiple celestial bodies within a star system.

To illustrate, we apply the logistic model to approximate global population dynamics between 1800 and 2100. For simplicity, we assume:
- $P_{max} = 12 \times 10^9$ (consistent with UN medium-fertility projections and ecological limits for Earth),
- $P_0 = 1 \times 10^9$ at $t_0 = 1800$,
- Intrinsic growth rate $r \approx 0.015 \text{yr}^{-1}$ (1.5% annual, declining over time).

Using these assumptions, we estimate $P(t)$ and the projected population for key reference years (Table 8 and Figure 8) as:

Table 8: Estimated $P(t)$ and the projected population

| Year | Estimated Global Population P(t) | Historical/Projected Actual Population | Notes |
|---|---|---|---|
| **1800** | ~1.0 × 10⁹ | ~1.0 × 10⁹ | Baseline pre-industrial era |
| **1900** | ~1.6 × 10⁹ | ~1.6 × 10⁹ | Industrial revolution expansion |
| **2000** | ~6.0 × 10⁹ | ~6.1 × 10⁹ | Modern era, demographic transition begins |
| **2100*** | ~10–11 × 10⁹ | ~10–11 × 10⁹ (UN projection) | Near carrying capacity, growth stabilizing |

*Projection assumes Earth-based carrying capacity. Expansion into extraterrestrial habitats would raise $P_{max}$ substantially, allowing further growth.*

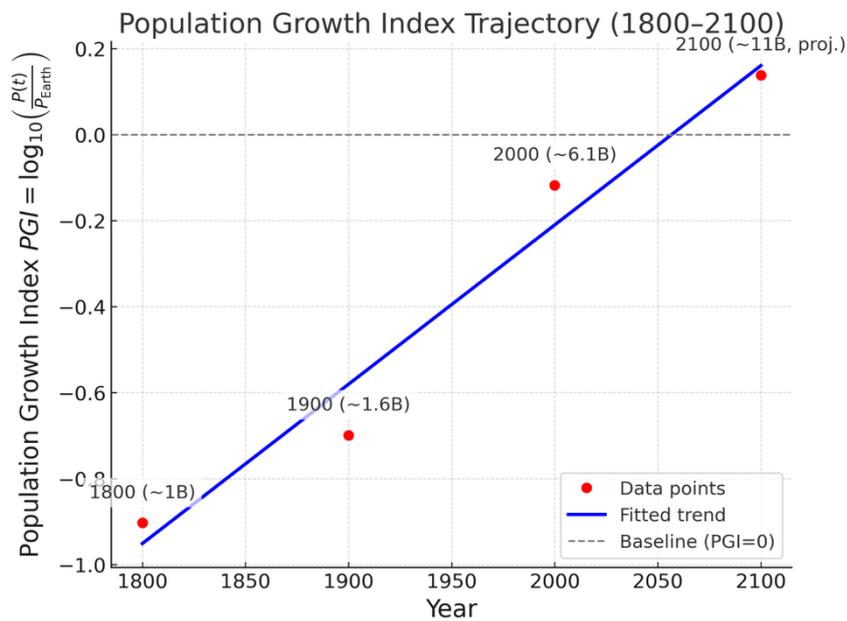

**Figure 8.** Historical and projected Population Growth Index (PGI) values from 1800 to 2100. Baseline $P_{Earth}$ corresponds to the approximate global population in 2023. Historical estimates (1800–2000) show sub-baseline PGI values, while projections for 2100 suggest PGI > 0, marking the transition to above-baseline population growth, potentially extended further by space colonization.

In scenarios involving interplanetary or interstellar expansion, a more flexible model such as a logarithmic or piecewise growth model may be preferable, reflecting continuous expansion and adaptation across diverse habitats and resource bases. In such frameworks, the carrying capacity dynamically evolves in response to technological innovations, spatial distribution, and resource extraction capabilities, enabling sustained population growth across vast interstellar distances.

### 2.4.4. The Role of Technology in Population Management

Technological innovation is fundamental to effectively managing the challenges associated with population growth. Advances in healthcare, biotechnology, artificial intelligence (AI), and robotics will significantly shape future population dynamics by improving life expectancy, reducing mortality, and elevating the overall quality of life. Breakthroughs in areas such as genetic engineering, personalized medicine, disease prevention, and regenerative therapies will allow civilizations to sustain larger, healthier populations, fueling further societal expansion and advancement.

AI and automation will play indispensable roles in efficiently managing the logistical complexities of large and increasingly dense populations, whether in terrestrial megacities or in extraterrestrial settlements. AI-driven systems could optimize the distribution of critical resources, streamline energy management, and enhance infrastructure maintenance, ensuring that population growth proceeds sustainably without overwhelming ecological or infrastructural limits.

Furthermore, technologies enabling interplanetary colonization—including advanced propulsion systems, space elevators, sustainable closed-loop life-support systems, and autonomous habitat construction—will be critical in extending the spatial boundaries for population growth. By facilitating the colonization of new environments beyond Earth, these technologies ensure that population growth can occur sustainably, with resource availability scaling appropriately in tandem with advancements in infrastructure and resource management systems.

### 2.4.5. Measuring Population Growth Potential

To systematically quantify the capacity for population growth as civilizations evolve, we introduce the Population Growth Index (PGI), defined as follows:

$$PGI = log_{10}\left(\frac{P(t)}{P_{Earth}}\right) \qquad (11)$$

where $P(t)$ is the population size at time $t$, and $P_{Earth}$ represents the baseline population size of Earth in the current era. For reference, we adopt $P_{Earth} = 8 \times 10^9$ (the approximate global population in 2023) as the baseline normalization.

This logarithmic index provides a standardized metric for comparing population growth potential across civilizations and developmental stages. As civilizations approach Type I status, PGI is expected to rise, reflecting increases in energy availability, efficient resource utilization, and the expansion of populations into extraterrestrial habitats. Subsequent interplanetary growth will further elevate PGI, underscoring a civilization's ability to sustain and expand large populations across planetary and interplanetary scales.

To illustrate, we estimate PGI for selected years in Table 9 and Figure 9:

**Table 9: Estimated $P(t)$ and the resultant PGI**

| Year | Global Population $P(t)$ | $PGI = log_{10}(P(t)/P_{Earth})$ | Notes |
|---|---|---|---|
| 1800 | ~1 × 10⁹ | -0.90 | Pre-industrial, agrarian economies dominate |
| 1900 | ~1.6 × 10⁹ | -0.70 | Early industrialization, rising urbanization |
| 2000 | ~6.1 × 10⁹ | -0.12 | Modern era, rapid 20th-century growth |
| 2100* | ~11 × 10⁹ (projected) | +0.14 | UN medium-fertility scenario, early space colonization possible |

*Projection based on UN demographic forecasts, with the possibility of extension if interplanetary habitats contribute additional capacity.*

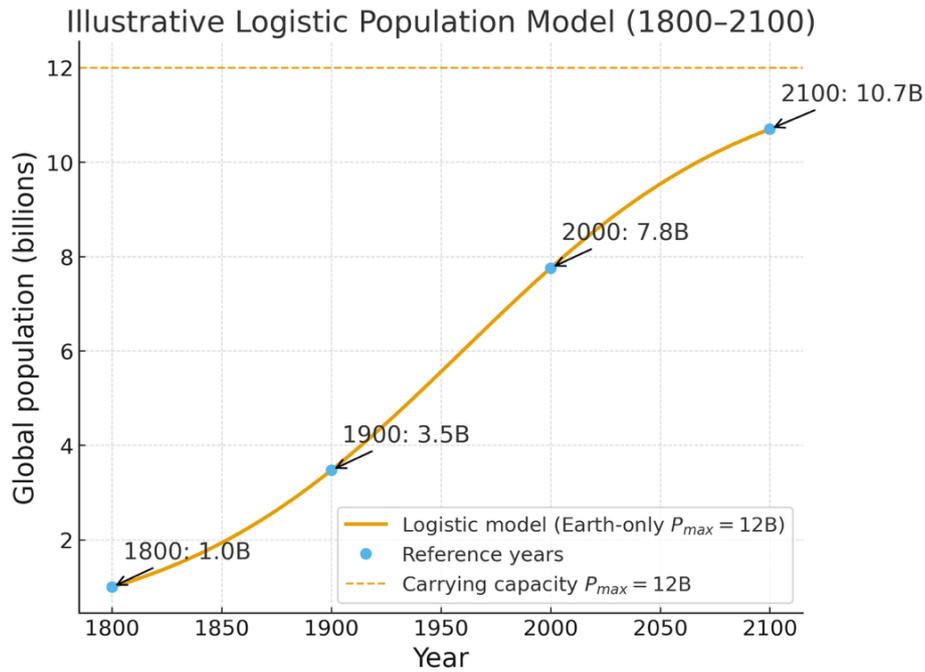

**Figure 9.** Illustrative logistic model of global population, 1800–2100, with $P_{max} = 12B$ (Earth-only carrying capacity), $P_0 = 1B$ at $t_0 = 1800$, and $r = 0.015\text{yr}^{-1}$. The curve approaches $P_{max}$ by ~2100. Off-world colonization would effectively increase $P_{max}$, shifting the asymptote upward.

### 2.4.6. Population Growth and Civilizational Sustainability

As civilizations expand in population size and spatial distribution, ensuring the sustainability of this growth becomes paramount. Long-term civilizational survival hinges upon the careful balancing of population demands with available resources, energy consumption, environmental conservation, and equitable societal structures. Sustainable population growth strategies necessitate the integration of advanced technological solutions with comprehensive governance systems that emphasize resource efficiency, equitable distribution, environmental stewardship, and adaptive policymaking.

For civilizations aspiring toward Type I and eventually Type II statuses, the central challenge will be maintaining a dynamic equilibrium between growth, resource utilization, and ecological preservation. Advanced, AI-guided governance frameworks that leverage comprehensive data analytics will be crucial for overseeing and managing sustainable population expansion on Earth and across interplanetary settlements. These systems will enable informed decision-making, predictive modeling of resource usage, and adaptive management of ecosystems and infrastructures, ensuring that population growth enhances rather than threatens long-term civilizational stability.

In summary, population growth potential is integral to civilizational advancement along the Kardashev scale. As populations grow and spread beyond planetary boundaries, they catalyze innovation, increase societal complexity, and drive the creation of sophisticated infrastructures. Managing this growth sustainably—balancing resource availability, technological progress, and environmental responsibility—will be critical for civilizations striving toward Type I and ultimately Type II statuses, securing their long-term viability and prosperity in the cosmos.

### 3. Unified Analytical Framework and Time Scale

Building upon our in-depth analyses across the Energy Scale, Information Scale, Large-Scale Construction Mass, and Potential for Population Growth, we now propose a comprehensive and unified analytical framework for assessing humanity's trajectory toward a Type II civilization. This integrated framework synthesizes individual components, allowing for a holistic evaluation of civilization advancement and providing robust projections across multiple critical dimensions.

**3.1 Integrated Analytical Model**

Our unified analytical framework systematically integrates the four primary metrics—Energy Scale, Information Scale, Large-Scale Construction Mass, and Population Growth Index—into a comprehensive quantitative model. This integrative approach acknowledges the interdependent nature of these scales; each influencing and being influenced by the others. Specifically:

- Energy availability constrains computational capacity and infrastructure scale.
- Information processing capability governs technological innovation, optimizing energy use, population management, and infrastructure efficiency.
- Large-scale construction mass represents the physical manifestation of energy utilization and directly enables the expansion and sustainability of both population and computational infrastructure.
- Population dynamics drive energy and resource demands, technological innovation, and infrastructural growth.

By capturing these interdependencies, we derive a unified Civilization Development Index (CDI), mathematically expressed as a weighted logarithmic composite of the individual indices:

$$CDI = \alpha \cdot K_{(P|Y)} + \beta \cdot \log_{10}\left(\frac{Compute\ Capacity\ (bits/s)}{Current\ Human\ Compute\ Capacity\ (bits/s)}\right) + \gamma \cdot C_M + \delta \cdot PGI \quad (12)$$

where:

- $K_{(P|Y)}$ represents the Kardashev Energy Scale metric (Eq. 5),
- Compute Capacity (bits/s) is derived from structural and energy considerations for optimal computing substrates (Eq. 7), normalized against current human computational capacity,
- $C_M$ is the logarithmic Construction Mass Index (Eq. 9),
- $PGI$ is the logarithmic Population Growth Index (Eq. 11), and
- $\alpha, \beta, \gamma, \delta$ are dimensionless weighting coefficients reflecting the relative influence of each parameter on civilization advancement.

The weighting coefficients can be dynamically adjusted based on future empirical data and theoretical insights into their evolving importance. By construction, CDI ≈ 1.0 corresponds to Type I civilization status, while CDI ≈ 2.0 corresponds to Type II, ensuring consistency with the Kardashev framework and the temporal projections discussed in Section 3.2.

**3.2 Temporal Projection and Milestones**

Utilizing current data trends, historical benchmarks, and projected technological advancements, we construct a detailed temporal framework that estimates humanity's progress along the integrated Civilization Development Index (CDI). Projections from our computational simulations, combined with historical extrapolation methods, indicate several key milestones. Our modeling approach complements prior simulation work on civilizational survival pathways and long-horizon risk/trajectory analysis [14].

To generate these projections, we adopted weighting coefficients for the CDI formula that reflect the relative importance of each scale at humanity's current stage of development:

- α = 0.35 for the Energy Scale, as energy availability underpins all technological, social, and infrastructural development.
- β = 0.25 for the Information Scale, reflecting the accelerating role of computation and data processing in driving innovation and societal organization.
- γ = 0.25 for the Large-Scale Construction Mass, given its critical role in enabling off-world infrastructure and megastructures.
- δ = 0.15 for the Population Growth Index, acknowledging its importance while recognizing that sustainable management is more critical than raw growth at advanced stages.

Sensitivity tests were performed with slightly varied coefficients, adjusting each weighting factor by ±0.05 (for example, the baseline α = 0.35 was tested at both 0.30 and 0.40, with analogous adjustments applied to β, γ, and δ). While exact milestone dates shift by decades to a century under these perturbations, the overall trajectory and ordering of milestones remain robust.

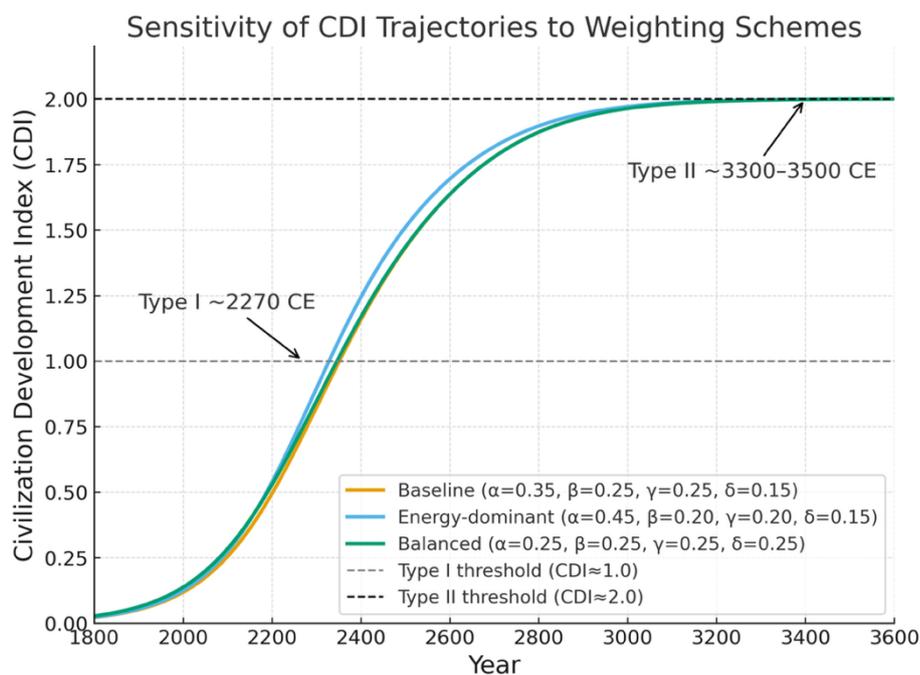

**Figure 10.** Sensitivity of the Civilization Development Index (CDI) trajectories to alternative weighting schemes for energy (α), information (β), construction mass (γ), and population (δ). The baseline scheme (α=0.35, β=0.25, γ=0.25, δ=0.15) highlights energy as the primary driver, while sensitivity tests with balanced and energy-dominant weightings show only modest shifts in timing. Type I status (CDI ≈ 1.0) is projected around 2270 CE, while Type II status (CDI ≈ 2.0) emerges between 3300–3500 CE under optimistic technological growth and resource utilization scenarios. This robustness underscores that milestone ordering remains stable despite variations in coefficient assignments.

**Milestones predicted under this weighting scheme:**

- Type I civilization status (CDI ~1.0) is likely to be achieved around 2271 CE, marked by complete planetary energy harnessing, global computational integration, substantial off-planet infrastructure, and sustainable global population management.
- Intermediate milestone (2300–2500 CE): robust computational infrastructure capable of processing on the order of $10^{46}$ bits per second, deployment of advanced off-world megastructures, and the establishment of sustainable interplanetary populations.

- Type II civilization status (CDI approaching 2.0): requiring fully operational stellar-scale infrastructures such as Dyson swarms, Matrioshka Brains, and stable interplanetary population distribution. Under optimistic growth rates and resource utilization scenarios, this may occur between 3200–3500 CE.

**4. Conclusion and Discussion**

Our unified analytical framework significantly extends and refines Kardashev's original scale by incorporating critical dimensions—energy utilization, information processing capacity, large-scale construction mass, and population dynamics. These dimensions are deeply interconnected, each facilitating and constraining the others. Energy availability underpins all aspects of civilizational progress, acting as the fundamental driver that enables advancements in information processing, infrastructure construction, and sustainable support for growing populations. Conversely, breakthroughs in computational technology and information processing foster innovation in energy extraction, efficiency, resource utilization strategies, and adaptive population management.

Large-scale construction mass represents a crucial yet often overlooked prerequisite for achieving stellar-scale energy-harvesting infrastructure. The advancement of autonomous robotics, additive manufacturing, nanotechnology, and materials science is essential for scaling construction projects from planetary to stellar realms. Without the ability to construct and maintain extensive megastructures such as Dyson Swarms or Matrioshka Brains, the immense energy resources of a star remain beyond practical reach.

Population growth serves as both a catalyst and a challenge. It provides the intellectual capital necessary for innovation, complexity, and sustained civilizational development. However, unchecked population expansion without corresponding advancements in technology and resource management risks ecological and infrastructural collapse. Thus, balancing growth with sustainability through advanced governance systems, informed by real-time data, robust predictive modeling, and adaptive artificial intelligence, becomes critical.

Strategically, this framework underscores several key interdependencies that humanity must manage to advance toward Type II civilization status. Firstly, energy remains both a fundamental constraint and an enabler. Targeted investment in advanced renewable energy technologies, nuclear fusion, and revolutionary energy-capturing infrastructures such as Dyson swarms will be crucial.

Secondly, enhancing computational capacity, bound by thermodynamic constraints as defined by Landauer's Principle and breakthroughs in materials science, must advance in step with energy availability. Effective integration of optimal computational substrates such as diamond, aluminum oxide, or titanium carbide can significantly maximize efficiency and scalability, although computational capacity will still be subject to the limits imposed by quantum mechanics.

Thirdly, achieving stellar-scale construction capabilities requires sustained technological progress in autonomous robotics and additive manufacturing. Development of self-replicating construction systems (e.g., Von Neumann machines) capable of harnessing extraterrestrial resources could dramatically accelerate human progression.

Fourthly, sustainable management of population growth is essential. Technological advancements in healthcare, biotechnology, artificial intelligence, and space colonization must be strategically managed to avoid resource scarcity and ecological imbalance. Ensuring carrying capacities expand in parallel with population growth through sustainable practices will be paramount.

Our quantitative projections suggest humanity could attain Type I civilization status around 2271 CE, assuming consistent progress in renewable and nuclear energy adoption, advanced computational capabilities, robust construction infrastructure, and sustainable population growth. Reaching Type II civilization status—projected between 2800 and 3200 CE—demands unprecedented global cooperation, visionary planning, and transformative technological innovation.

It is essential to recognize that projecting timelines inherently involves uncertainties, particularly concerning technological breakthroughs, global geopolitical dynamics, and unforeseen existential threats. Our model presumes continuous technological growth and uninterrupted societal advancement—conditions that require proactive management, strategic foresight, and global collaboration on scales that have, historically speaking, been only inconsistently demonstrated. These conclusions are consistent with our prior simulation-based analysis of long-horizon civilizational risk factors and survival pathways, which highlighted the amplifying role of governance, coordination, and sustained technological investment in shaping favorable trajectories [24].

Ultimately, humanity's journey toward Type II civilization status is not solely defined by engineering achievement or resource availability. It embodies a fundamental transformation in societal organization, consciousness, and our collective relationship with the cosmos. Achieving this evolutionary milestone will involve profound shifts in governance structures, ethical considerations, and a broader understanding of humanity's role within the universe. This comprehensive analytical framework thus serves as both a practical guide for strategic investments and a conceptual lens through which humanity can envision and pursue a sustainable and prosperous future in the cosmos.

Notes

1. Based on energy consumption, humanity is currently at Type ~0.72.
2. Given that some other large-brained species also exhibit consciousness (assuming this to be effectively equivalent to demonstrated self-awareness), such as dolphins and higher primates, "human" may be a bit too narrow.

**Acknowledgements**

This research was supported by the NASA-sponsored Jet Propulsion Laboratory, California Institute of Technology, and the Department of Computer Engineering, Sejong University. The authors thank Philip E. Rosen for his helpful comments and suggestions, which strengthened the manuscript.

**References**

**[1]** Gray, R. H. (2020). The Extended Kardashev Scale, *The Astronomical Journal*, 159, 228.

**[2]** Ćirković, M. M. (2015). Kardashev's classification at 50+: A fine vehicle with room for improvement. *Serbian Astronomical Journal*, vol. 191, pp. 1-15.

**[3]** Kardashev, N. S. (1964). Transmission of Information by Extraterrestrial Civilizations. *Soviet Astronomy*, 8, 217.

**[4]** Lemarchand, G. A. (1998). Detectability of Extraterrestrial Technological Activities. COS-ETI. http://www.coseti.org/lemarch1.htm

**[5]** Ćirković, M. M. (2018). The Great Silence: Science and Philosophy of Fermi's Paradox. Oxford University Press, New York.

**[6]** Manfredi, G. (2020). Exoplanets and Habitability. *Sociedade Astronômica Brasileira,* 31, no. 1, 21-22. https://sab-astro.org.br/wp-content/uploads/2020/04/GabrielManfredi.pdf

**[7]** Clark, P. (n.d.). Spectral Classification Notes. University College London. http://www.star.ucl.ac.uk/~pac/spectral_classification.html

**[8]** International Energy Agency (2022). World Energy Outlook 2022: An Updated Roadmap to Net Zero Emissions by 2050. https://www.iea.org/reports/world-energy-outlook-2022/an-updated-roadmap-to-net-zero-emissions-by-2050

**[9]** Statista (2025). Nuclear Power Market Outlook. https://www.statista.com/outlook/io/energy/nuclear-power/worldwide

**[10]** Jiang, J. H., Feng, F., Rosen, P. E., et al. (2022). Avoiding the Great Filter: Predicting the Timeline for Humanity to Reach Kardashev Type I Civilization. *Galaxies*, 10(3), 68. https://doi.org/10.3390/galaxies10030068

**[11]** De Weck, O. L. (2005). Technology Roadmapping and Development: A Quantitative Approach to the Management of Technology. Springer.

**[12]** AI Impacts (2015). Brain Performance in FLOPS. https://aiimpacts.org/brain-performance-in-flops

**[13]** TOP500.org (2025). TOP500 Supercomputer Sites. https://top500.org

**[14]** Bradbury, R. J. (1999). Matrioshka Brains. Online essay: https://gwern.net/doc/ai/scaling/hardware/1999-bradbury-matrioshkabrains.pdf

**[15]** Landauer, R. W. (1961). Irreversibility and heat generation in the computing process. *IBM Journal of Research and Development*, **5**(3), 183–191, doi:10.1147/rd.53.0183.

**[16]** An, K., Wu, H., Liu, P., Yang, Z., … & Li, C. (2024). Accurate comparison of the fracture toughness of ultra-thick polycrystalline diamond plates by ISO standard method. *Diamond and Related Materials, 149*, 111585.

**[17]** Prielipp, H., Knechtel, M., Claussen, N., Streiffer, S. K., Müllejans, H., Rühle, M., & Rödel, J. (1995). Strength and fracture toughness of aluminum/alumina composites with interpenetrating networks. *Materials Science & Engineering A*, **197**(1), 19-30.

**[18]** ScienceDirect. Titanium Carbide – An Overview. https://www.sciencedirect.com/topics/earth-and-planetary-sciences/titanium-carbide

**[19]** Panadyne Inc. (2025). *Aluminum Oxide Technical Data Sheet*. https://www.panadyne.com

**[20]** Exploding Topics (2025). How Much Data is Generated Each Day? https://explodingtopics.com/blog/data-generated-per-day

**[21]** Freitas, R. A., Jr. (1979). Xenology: An Introduction to the Scientific Study of Extraterrestrial Life, Intelligence, and Civilization (1st ed.). Xenology Research Institute. Chapter 14.3: "Alien Consciousness / Sentience Quotient."

**[22]** Ćirković, M. M. (2000). On the Inevitability and the Possible Structures of Supercivilizations. *Foundations of Physics*, Vol. 30(12).

**[23]** Bremermann, H. J. (1962). Optimization through evolution and recombination. In M. C. Yovits, G. T. Jacobi & G. D. Goldstein (Eds.), *Self-Organizing Systems* (pp. 93-106). Spartan Books, Washington, D.C.

**[24]** Jiang, J. H., et al. (2023). Avoiding the Great Filter: A Simulation of Important Factors for Human Survival. *J. Human. Soc. Sci.*, 6(1), 33-54.